\documentclass[12pt]{article}

\usepackage{a4wide}
\usepackage{amssymb}
\newtheorem{theorem}{Theorem}    % Specify Theorem
\newtheorem{lemma}{Lemma}    % Specify Theorem
\newtheorem{result}{Result}    % Specify Result
\newtheorem{claim}{Claim}      % Specify Claim
\newcommand{\qed}{\hfill{$\rule{6pt}{6pt}$}} %Box at end of proof
\newenvironment{proof}{\noindent{\bf Proof}:}{\qed}

\newcommand{\reals}{\mathbb R}
\newcommand{\defeq}{\stackrel{\Delta}{=}}
\newcommand{\ceil}[1]{\left\lceil #1 \right\rceil} 

\newcommand{\ket}[1]{| \phi_{#1} \rangle}
\newcommand{\kettle}[1]{| #1 \rangle}
\newcommand{\braket}[2]
  {\langle \phi_{#1} | W_S^\dagger W_T | \phi_{#2} \rangle}
\newcommand{\inprod}[2]{\langle #1 | #2 \rangle}

\title{{\bf The quantum complexity of set membership}\thanks{
A preliminary version of this paper appeared in the {\em Proceedings
of the 41st Annual IEEE Symposium on Foundations of Computer Science},
pages 554--562, 2000.}
}

\author{
Jaikumar~Radhakrishnan\thanks{ 
School of Technology and Computer Science, 
Tata Institute of Fundamental Research, Mumbai 400005, India.
Email: {\sf jaikumar@tcs.tifr.res.in}.} 
\and
Pranab~Sen\thanks{
Laboratoire de Recherche en Informatique, 
Universit\'{e} de Paris-Sud, 91405 Orsay, France.
Email: {\sf pranab@lri.fr}.
Most of this work was done while the author was a
graduate student at the Tata Institute of Fundamental Research.} 
\and
S.~Venkatesh\thanks{
Center for Discrete Mathematics and Theoretical Computer Science (DIMACS), 
Rutgers University, 
Piscataway, NJ 08854, USA. 
Email: {\sf venkat@dimacs.rutgers.edu}. 
Most of this work was done while the author was a
graduate student at the Tata Institute of Fundamental Research.} 
}

\date{}

\begin{document}

\maketitle

\begin{abstract}
We study the {\em quantum complexity} of the static set membership
problem: given a subset $S$ ($|S|\leq n$) of a universe of size $m$
($\gg n$), store it as a table, $T:\{0,1\}^r \rightarrow \{0,1\}$, of
bits so that queries of the form `Is $x$ in $S$?' can be answered. The
goal is to use a small table and yet answer queries using few
bit probes. This problem was considered recently by Buhrman,
Miltersen, Radhakrishnan and Venkatesh~\cite{buhrman:bitprobe}, who 
showed lower and upper bounds for this problem in the classical
deterministic and randomised models. In this paper, we formulate this
problem in the ``quantum bit probe model''.
We assume that access to the table $T$ is provided by means of
a black box
(oracle) unitary transform  $O_T$ that takes the basis state
$\kettle{y,b}$ to the basis state $\kettle{y,b\oplus T(y)}$. The query
algorithm is allowed to apply $O_T$ on any superposition of basis
states.

We show tradeoff results
between space (defined as $2^r$) and number of probes (oracle calls)
in this model.  Our results show
that the lower bounds shown in~\cite{buhrman:bitprobe} 
for the classical model
also hold (with minor differences) in the quantum bit probe
model. These bounds almost match the classical upper bounds. Our lower
bounds are proved using linear algebraic arguments.
\end{abstract}

\noindent {\bf Keywords:} Data structures, set membership,
bit probe model, quantum black box model, linear 
algebraic methods, lower bounds, space-time tradeoffs.

\section{Introduction}

In this paper we study the {\em static membership} problem: Given a
subset $S$ ($|S| \leq n$) of a universe of size $m$ ($\gg n$), 
store it efficiently and succinctly so that 
queries of the form ``Is
$x$ in $S$?'' can be answered quickly. This fundamental data structure
problem has been studied earlier in various settings (e.g. by Minsky and
Papert~\cite{minsky:perceptrons}, Yao~\cite{yao:tablesort}, 
Fredman, Koml\'{o}s and
Szemer\'{e}di~\cite{fredman:setmemb}, Pagh~\cite{pagh:setmemb}).
Most of these results were in the classical
deterministic {\em cell probe model}. Recently, this 
problem was considered
by Buhrman, Miltersen, Radhakrishnan and 
Venkatesh~\cite{buhrman:bitprobe} in the
classical {\em bit probe model}, which was introduced in 
\cite{minsky:perceptrons}; they studied
tradeoffs between storage space and number of probes in the classical
deterministic case, and also showed lower and upper bounds for the
storage space when the
query algorithm was randomised and made just {\em one} bit probe.  
In the classical 
bit probe model the storage scheme is deterministic and
stores the given set as a string
of bits. The query scheme is either deterministic or randomised and
answers membership queries probing
only one bit of the string at a time.

In this paper, we allow the query algorithm to perform quantum search
on the table of bits representing the 
stored subset $S$. The storage scheme which encodes $S$
a table of bits, $T:\{0,1\}^r \rightarrow \{0,1\}$, continues to 
be classical deterministic. 
To formalise this, we define the {\em quantum
bit probe model}. The table of $2^r$ bits, $T$,
is modelled using an oracle unitary transformation $O_T$
that takes the basis state $\kettle{y,b}$ to $\kettle{y,b\oplus T(y)}$.
Basically, the table of bits is accessed by the query algorithm as
in the well-studied {\em quantum black box model} 
(see e.g.~\cite{beals:quantpoly}).
If the inputs to the oracle $O_T$ are restricted to basis states, 
$O_T$ reduces to (the reversible version of) the classical table that
stores one bit for each address in $\{0,1\}^r$. We, therefore, define
the {\em space} used by the quantum bit probe scheme to be $2^r$.

The main point of departure from the classical model, is in the query
algorithm. We allow the algorithm to feed a superposition of basis
states to the oracle.  Each use of the oracle counts as one probe of
the table, even if a superposition is supplied to the oracle, and the
output depends on the value of several bits of the underlying table
$T$. In the preparation of this superposition and in the processing
of the output returned by the oracle, we allow arbitrary unitary
transformations.  It is known that this form of access often leads to
significant improvements over classical algorithms for several problems
(e.g. Grover's algorithm~\cite{grover:search} for 
searching an unordered database).

Previously, the number of probes to the black box as a complexity
measure had been studied in the 
quantum setting (e.g. \cite{bennett:strengths,
beals:quantpoly, ambainis:quantlb, grover:search}). Both 
lower bounds and upper
bounds for various problems were proved. 
The main contribution of this paper is the study of tradeoffs between
storage space and number of probes for a static data structure problem
in the quantum setting.
For the set
membership problem, we show that several limitations of
classical computation (shown in \cite{buhrman:bitprobe}) 
continue to persist even
if quantum query algorithms are allowed. This is surprising, because
for the (superficially) similar problem of searching an unordered 
database, quantum computation helps. 

Our tradeoffs between storage space and the number of quantum probes
are proved using linear algebraic arguments. 
Roughly speaking, we lower and upper bound the dimension of a set
of unitary operators arising from the quantum query algorithm. The
lower bound on the dimension arises from the `correctness requirements'
of the quantum algorithm. The upper bound on the dimension arises
from limitations on the storage space and number of probes. 
By playing the lower and upper
bounds against each other, we get the desired tradeoffs. 
To the best of our knowledge, this is the first time that linear
algebraic arguments have been used to prove lower
bounds for data structure problems, classical or quantum. Counting
of dimensions has been previously used in quantum computing (see
e.g.~\cite{ambainis:sampling, buhrman:ccpoly}), but in quite
different contexts and ways. Linear algebraic arguments similar
to ours have been heavily used in combinatorics. For a 
delightful introduction, see the book
by Babai and Frankl~\cite{babai:linalg}.

\subsection{Our results}

\paragraph{The  exact quantum model:} 

Buhrman {\em et al.}~\cite{buhrman:bitprobe} have shown, 
for classical deterministic query algorithms, that
any $(s,t)$-scheme (which uses space $s$ and $t$ bit probes) satisfies
${m \choose n} \leq {s \choose nt}2^{nt}$. We show a
stronger (!) tradeoff result in the quantum bit probe model.
\begin{quote}
\begin{result}
\label{result:exactquant}
Suppose there exists a scheme for storing subsets $S$ of
size at most $n$ from a universe {\bf U} of size $m$ that uses 
$s$ bits of storage and answers membership queries, with
zero error probability, with $t$ quantum probes.
Then,
\[ \sum^{n}_{i=0} {m \choose i} \leq \sum^{nt}_{i=0}{s \choose i}\]
\end{result}
\end{quote}
This has two immediate consequences. First, by setting $t=1$, we see
that if only one probe is allowed, then $m$ bits of storage are
necessary. (In \cite{buhrman:bitprobe}, for the classical 
model, this was
justified using an ad hoc argument.) Thus, the classical deterministic
{\em bit vector} scheme that stores the characteristic vector of
the set $S$ and answers membership queries using one bit probe, is
optimal even up to quantum. Second, it 
follows (see \cite{buhrman:bitprobe} for
details) that the
classical deterministic scheme of Fredman, Koml\'{o}s and
Szemer\'{e}di~\cite{fredman:setmemb}, which 
uses $O(n \log m)$ bits of storage and
answers membership queries using $O(\log m)$ bit 
probes, is optimal even up to
quantum --- quantum schemes that use $O(n\log m)$ bits of
storage must make $\Omega(\log m)$ probes if 
$n \leq m^{1 - \Omega(1)}$.
Recently, Pagh~\cite{pagh:setmemb} has shown classical 
deterministic schemes
using the information-theoretic minimum space $O(n \log (m/n))$ and
making $O(\log (m/n))$ bit probes, which is optimal even up to quantum,
by the above result.
For $t$ between $1$ and $O(\log (m/n))$, 
Buhrman {\em et al.}~\cite{buhrman:bitprobe}
have given classical deterministic schemes making $t$ bit probes,
which $O(m^{3/t} n \log m)$ bits of storage. A lower bound of
$\Omega(n t (m/n)^{1/t})$ for storage space, for suitable values 
of the various
parameters, follows from Result~\ref{result:exactquant}. 
Thus, if we only care about
space up to a polynomial, classical deterministic schemes that
make $t$ bit probes for $t$ between $1$ and $O(\log (m/n))$, and which
use storage space almost matching the exact quantum lower bounds, exist.

Interestingly, the above theorem holds even in the presence of errors,
provided the error is restricted to positive instances, that is the
query algorithm sometimes (with probability $< 1$) returns the answer
`No' for a query $x$ that is actually in the set $S$.
This was not observed earlier even in the
classical model, although one can easily modify the proof of the
tradeoff result in \cite{buhrman:bitprobe} to give this. 

\paragraph{The $\epsilon$-error model:}
In the classical model, there exists a 
scheme for storing subsets of size at most $n$ from a
universe of size $m$ that answers membership queries, with two-sided
error at most $\epsilon < 1/16$, using just 
{\em one} bit probe, and using
storage space $O(\frac{n \log m}{\epsilon^2})$. Also, any such one
probe scheme making two-sided error at most $\epsilon$ must
use space $\Omega(\frac{n \log m}{\epsilon \log (1/\epsilon)})$. Both
the upper bound and the lower bound have been proved in 
\cite{buhrman:bitprobe}. By two-sided error, we mean 
that the query algorithm
can make an error for both positive instances (the query element
is a member of the stored set), as well as negative instances (the
query element is not a member of the stored set).
Since different sets must be represented by different
tables, every scheme, no matter how many probes the query algorithm
is allowed, must use $\Omega( n \log (m/n))$ bits of storage, 
even in the bounded two-sided error
quantum model. However, one might ask if the dependence of space on
$\epsilon$ is significantly better in the quantum probe model. We
show the following lower bound which implies that a quantum scheme
needs significantly more than the information-theoretic optimal
space if sub-constant error probabilities are desired.
\begin{quote}
\begin{result}
\label{result:twosidedquantone}
Let $n/m < \epsilon < 1/8$.
Suppose there is a scheme with two-sided error $\epsilon$
which stores subsets of size at most $n$ from a universe of size
$m$ and
answers membership queries, with two-sided error at most $\epsilon$,
using one quantum probe. It must
use space 
\[ s=\Omega \left(\frac{n\log (m/n)}{\epsilon^{1/6}\log (1/\epsilon)}
            \right)
\]
\end{result}
\end{quote}

The method used to prove this result can be generalised to algorithms
that use more probes than one.
\begin{quote}
{\bf Result~\ref{result:twosidedquantone}'} 
{\it 
For any $p \ge 1$ and $n/m < \epsilon < 2^{-3p}$,
suppose there is a scheme 
which stores subsets of size at most $n$ from a universe of size
$m$ and
answers membership queries, with two-sided error at most $\epsilon$,
using $p$ quantum probes. Define $\delta \defeq \epsilon^{1/p}$.
It must use space 
\[ s=
 \Omega\left(\frac{n\log (m/n)}
                  {\delta^{1/6} \log (1/\delta)}
       \right)\]
\/}
\end{quote}

Such a tradeoff between space and error probability for multiple
probes was not known 
earlier, even in the classical randomised model.
We note that for $p$ bit probes, an upper bound of 
$O(\frac{n \log m}{\epsilon^{4/p}})$ on the storage space,
for $\epsilon < 2^{-p}$, follows by 
taking the storage scheme of
\cite{buhrman:bitprobe} for error 
probability $\frac{\epsilon^{2/p}}{4}$, 
and repeating the (classical randomised)
single probe query scheme $p$ times. This diminishes the
probability of error to $\epsilon$.
Thus, our lower bounds for two-sided error quantum
schemes roughly match the two-sided error classical randomised 
upper bounds.

The results described above are inspired by similar results proved
earlier in \cite{buhrman:bitprobe} in 
the classical model. However, the methods
used for classical models, which were based on combinatorial arguments
involving set systems, seem to be powerless in giving the results in
the quantum model. Our results are based on linear algebraic
arguments, involving counting the dimensions of spaces of
various operators that
arise in the quantum query algorithm. 

\paragraph{Bounds for classical models:} As stated above, 
Result~\ref{result:exactquant} is 
stronger than what was known earlier, even in the
classical deterministic model. 
One might wonder if this stronger result is somehow
easier to prove in the classical deterministic model. We 
show that the linear
algebraic techniques used in the proof of Result~\ref{result:exactquant}
can be considerably simplified when we assume the classical 
deterministic model, and give the same inequality as stated in 
Result~\ref{result:exactquant}. 
Also, one can easily modify the proof to yield the same tradeoff
for randomised query schemes where the error is restricted to positive
instances (i.e. when the query element is a member of the stored set).

The proof in \cite{buhrman:bitprobe} of the space lower bound for a
classical  two-sided $\epsilon$-error randomised query scheme, namely 
$s = \Omega( \frac{n \log m}{\epsilon \log (1/\epsilon)} )$, 
involved some tricky use of  both upper and lower bounds for
$r$-cover-free families shown by Nisan 
and Wigderson~\cite{nisan:hardness},
Erd\H{o}s, Frankl and F\"{u}redi~\cite{erdos:set}, 
Dyachkov and Rykov~\cite{dyachkov:set}.  The proof of
the analogous bound (Result~\ref{result:twosidedquantone}) in 
the quantum model completely avoids
these. In fact, we give a proof of a (slightly weaker)
lower bound $s =
\Omega( \frac{n \log m}{\epsilon^{2/5} \log (1/\epsilon)} )$ 
in the classical
randomised model by adapting the ideas used in the proof of 
Result~\ref{result:twosidedquantone}
to the classical setting.  We first diminish the error 
probability of the one-probe query algorithm by repetition and then
we can, by fixing the random coin tosses, make it a deterministic
query algorithm which however, uses more than one probe. We then apply
our (classical) deterministic space-time tradeoff equation 
to complete the proof. This approach
is inspired by our proof of Result~\ref{result:twosidedquantone}.
Besides being simpler, this proof has the
advantage that it generalises readily to more than one probe.
No such result was known earlier in the classical setting.
\begin{quote}
\begin{result}
\label{result:classicaltwosidedp}
Let $p \ge 1$, $18^{-p} > \epsilon > 1 / m^{1/3}$ and
$m^{1/3} > 18n$.
Define $\delta \defeq \epsilon^{1/p}$.
Any classical scheme which stores subsets
of size $n$ from a universe of size $m$ and 
answers membership queries, with two-sided error at most $\epsilon$,
using at most $p$ bit probes must use space 
\begin{displaymath}
\Omega \left(\frac{n \log m}{\delta^{2/5} \log (1 / \delta)}
        \right)
\end{displaymath}
\end{result}
\end{quote}

\subsection{Organisation of the paper}
In the next section, we describe our quantum bit probe model formally
and give the framework of our proofs.  Detailed proofs of all our
results (for both quantum and classical models) appear in
Section~\ref{sec:proofs}. We conclude in Section~\ref{sec:conclusion} 
with some open problems.

\section{Definitions and notations}
\label{sec:defnot}
In this section we first describe our {\it quantum bit probe model\/}
and then give some formal definitions and notations which will
be used in the proofs of the theorems.
\subsection{The model}
Our model is a quantum analogue of the classical bit probe model which
has been extensively used in the past to study data structure
problems (see e.g.~\cite{minsky:perceptrons,
miltersen:bitprobe, buhrman:bitprobe, pagh:setmemb}). 

A static data structure problem is
a function $f : D \times Q \rightarrow \{0,1\}$, where $D$ is a finite
set called the set of data to be stored and $Q$ is a finite set
called the set of queries. A classical $(s,t)$-bit probe scheme 
for a static data
structure problem consists of a deterministic storage scheme which 
stores the given
data $d$ as a bit string of length $s$, and a query scheme which given
a query $q$ makes at most $t$ bit probes to the stored string and 
computes
$f(d,q)$. The query scheme can be either deterministic or 
randomised. For more details about the classical model, see 
\cite{buhrman:bitprobe}. 

A quantum $(s,t)$-bit probe scheme for a static data structure problem
has two components: a classical deterministic
storage scheme that stores the data $d$ using
$s$ bits, and a quantum query scheme that answers queries by 
`quantumly probing a bit at a time' at most $t$ times.

\paragraph{The storage scheme:} For the set membership problem, the
data to be stored is a subset $S$ of a universe ${\mathbf U}$ 
($|S|\leq n$,
$|{\mathbf U}| =  m$). Let $x(S) \in \{0,1\}^s$ be the bit string 
that is
stored by the storage scheme for recording $S$. The storage scheme
is classical deterministic. The difference now, is that this 
bit string is made available to the
query algorithm in the form of an oracle unitary transform $O_S$. To
define $O_S$ formally, we represent the basis states of the
quantum query circuit as
$\kettle{j,b,z}$, where $j \in [s]$ is a binary string of
length $\log s$ (`address qubits'), $b$ is a single bit (`data qubit'),
and $z$ is a binary string of
some fixed length (`work qubits'). 
Let $x(S)_j$ be the bit stored at the $j$th location in the string
$x(S)$. The action of $O_S$ on a basis state is
described below.
\begin{displaymath}
O_S: \kettle{j,b,z} \mapsto (-1)^{b \cdot x(S)_j} \kettle{j,b,z}
\end{displaymath}

\smallskip
\noindent {\bf Remark:} The oracle described in the introduction mapped
$\kettle{j,b,z}$ to $\kettle{j,b\oplus x(S)_j,z}$. It is
known that the oracle $O_S$ defined
above is equivalent in power to this oracle. 

Thus, information about the string $x(S)$ appears in the
phase of the basis states in the output, and $O_S$ is represented by a
diagonal matrix (in the standard computational basis): the $i$th
diagonal entry, where $i \equiv \kettle{j,b,z}$, is 
\begin{displaymath}
(O_S)_{i,i} = (-1)^{b \cdot x(S)_j}
\end{displaymath}
For $T \subseteq [s]$ and $x \in \{0,1\}^s$, define
$[x]_T \defeq \sum_{i\in T} x_i \mbox{\ (mod 2)}$. In
particular, $[x]_{\emptyset}=0$. 
Thus, $(O_S)_{i,i} = (-1)^{[x(S)]_{l_i}}$, where
$l_i$ is some subset of $[s]$ of size $1$ (when
$b=1$, $l_i=\{j\}$) or $0$ (when $b=0$, $l_i=\emptyset$).

\smallskip
\noindent {\bf Remark:} Our model for storage does not permit $O_S$ to
be any arbitrary unitary transformation. However, this restricted form
of the oracle is closer to the way bits are accessed in the classical
case. Moreover, in most previous works, storage has been modelled using
such an oracle (see e.g. \cite{grover:search, bennett:strengths,
beals:quantpoly, ambainis:quantlb}).

\paragraph{The query scheme:} 
Suppose a subset $S \subseteq {\bf U}, |S| \leq n$, has been stored
and $x(S) \in \{0,1\}^{s}$ is the corresponding bit string.
A quantum query scheme with $t$ probes is just a sequence of
unitary transformations
\begin{displaymath}
U_{0} \rightarrow O_{S} \rightarrow U_{1} \rightarrow O_{S}
      \rightarrow \ldots U_{t-1} 
      \rightarrow O_{S} \rightarrow U_{t}
\end{displaymath}
where $U_{j}$'s are arbitrary unitary transformations that do not
depend on the set $S$ stored. For a query $q \in {\bf U}$, the
computation starts in an observational basis state 
$\kettle{q}\kettle{0}$, where we
assume that the ancilla qubits are initially in the basis state 
$\kettle{0}$. Then we apply the operators $U_{0}, O_{S}, \ldots,
O_{S}, U_{t}$ and measure the final state. The result of the query
is the rightmost bit of the state obtained by the measurement.
The query scheme can be exact or have error; the error
can be one-sided or two-sided. When the query scheme is exact, the
measurement of the final state gives the correct answer with
probability $1$. If one-sided error $\epsilon$ is allowed, 
the measurement
produces a $0$ with probability $1$ when the answer is $0$, but when
the answer is $1$, is required to produce a $1$ with probability only
at least $1-\epsilon$. If two-sided error $\epsilon$ is allowed, the 
answer can be wrong,
with probability at most $\epsilon$, for both positive and negative
instances. 

\paragraph{The framework for the proofs:}
We now describe the general framework in which the various proofs
are presented and also give some definitions and notations which 
will be used throughout the paper.

For a query $q \in {\mathbf U}$, define 
$\ket{q} \defeq \kettle{q}\kettle{0}$. The 
set of vectors $\ket{q}, q \in {\bf U}$ form an orthogonal system
of vectors. They are independent of the set $S$ stored.

Define two Hilbert spaces, $A_{0}$ and $A_{1}$, where $A_{i}$ is the
space of all state vectors that can be spanned by basis states having
an $i$ at the rightmost bit (i.e. if the state vector lies in $A_{i}$,
then on measuring the rightmost bit at the output, one gets $i$ with
probability $1$). Then the entire state space $V$ decomposes as an
orthogonal direct sum of the spaces $A_{0}, A_{1}$.

Define the unitary transformations 
$\left\{W_{S}\right\}_{S \subseteq {\bf U}, |S| \leq n}$ as follows.
\begin{displaymath}
W_{S} \defeq 
   U_{t}O_{S}U_{t-1}O_{S}U_{t-2}O_{S} \cdots U_{2}O_{S}U_{1}O_{S}U_{0}
\end{displaymath}

Thus when a set $S$ is stored, in the exact quantum case,
$W_{S}\ket{i}, i \in S$ lie in $A_{1}$, and
$W_{S}\ket{i}, i \not\in S$ lie in $A_{0}$.
In the one-sided $\epsilon$-error case, 
$W_{S}\ket{i}, i \not\in S$ lie in
$A_{0}$, but $W_{S}\ket{i}, i \in S$ may not lie entirely in $A_{1}$,
but in fact may have a projection on $A_{0}$ of length at most
$\sqrt{\epsilon}$.
In the two-sided $\epsilon$-error case, $W_{S}$
 has to send the vector $\ket{i}$ ``approximately'' to the correct 
space, i.e. the projection of $W_{S} \ket{i}$ on the correct space
is of length more than $\sqrt{1 - \epsilon}$.

\paragraph{Notation:}
In the proofs we have to 
take tensor products
of vectors and matrices. For any vector $v$ or matrix
$M$, we have the following notation,
\begin{displaymath}
v^{\otimes t} \defeq \underbrace{v \otimes \cdots \otimes v}_
                         {t \ {\rm times}}
\end{displaymath}
\begin{displaymath}
M^{\otimes t} \defeq \underbrace{M \otimes \cdots \otimes M}_
                         {t \ {\rm times}}
\end{displaymath}
We note that since the entire state space $V$ is the orthogonal
direct sum of $A_{0}$ and $A_{1}$, 
\begin{displaymath}
V^{\otimes t} = A_{0}^{\otimes t} \oplus 
                (A_{0}^{\otimes t-1} \otimes A_{1}) \oplus \cdots
		\oplus A_{1}^{\otimes t}
\end{displaymath}
and the $2^{t}$ vector spaces in the above direct sum are pairwise
orthogonal.

Below, $A \bigtriangleup B$ stands for the symmetric difference of sets
$A$ and $B$; $M^\dagger$ stands for the conjugate transpose of the
matrix $M$.

\section{Proofs of theorems}
\label{sec:proofs}

\subsection{Quantum schemes}
We first prove our space v/s probes tradeoff result for exact
quantum schemes.

\begin{theorem}
\label{thm:exactquant}
Suppose there exists a scheme for
storing subsets $S$ of
size at most $n$ from a universe ${\bf U}$ of size $m$
that uses $s$ bits of
storage and answers membership queries, with zero error probability,
with $t$ quantum probes.
Then,
\[ \sum^{n}_{i=0} {m \choose i} \leq \sum^{nt}_{i=0}{s \choose i}\]
\end{theorem}
\begin{proof}
We use the notation of Section~\ref{sec:defnot}.
For any subset $S \subseteq {\bf U}$, $|S| \leq n$, let us define
\[ W_S \defeq U_tO_SU_{t-1}O_{S}U_{t-2}O_S \cdots U_2O_SU_1O_SU_0 \]
\begin{claim}
$\{W_S^{\otimes n}\}_{S \in {{\bf U} \choose \leq n}}$ are 
linearly independent.
\end{claim}
\begin{proof}
Suppose there is a nontrivial linear combination
\[\sum_{S \in {{\bf U} \choose \leq n}} \alpha_S W_S^{\otimes n} = 0\]
Let $T$ be a set of largest cardinality such that $\alpha_T \neq 0$
and let $T=\{i_1,\ldots,i_k\}$, $k \leq n$. We define a vector
\[ \ket{T} \defeq \ket{{i_1}}^{\otimes (n-k+1)} \otimes \ket{{i_2}}
        \otimes \cdots \otimes \ket{{i_k}}\]
Applying $\ket{T}$ to the linear combination above, we have
\begin{equation}
\label{eq:exactlin}
\sum_{S \in {{\bf U} \choose \leq k}, S \neq T} 
 \alpha_S W_S^{\otimes n}\ket{T} + \alpha_T W_T^{\otimes n}\ket{T} = 0
\end{equation}
For any set $S$, 
\[
W_S^{\otimes n}\ket{T} = (W_S\ket{{i_1}})^{\otimes (n-k+1)} 
                           \otimes W_S\ket{{i_2}}
                           \otimes \cdots \otimes W_S\ket{{i_k}}
\]
\begin{itemize}
\item
If $S=T$, $W_T\ket{{i_l}} \in A_1$ for all $l$, $1 \leq l \leq k$
Hence $W_T^{\otimes n}\ket{T} \in A_1^{\otimes n}$.
\item
If $S \neq T$, there exists an element $i_j$ in $T-S$ (by 
choice of $T$).
$W_S\ket{{i_j}} \in A_0$. Hence 
$W_S^{\otimes n}\ket{T} \not \in A_1^{\otimes n}$. 
In fact, $W_S^{\otimes n}\ket{T}$ is orthogonal to $A_1^{\otimes n}$.
\end{itemize}

Hence, in the above linear combination (equation~\ref{eq:exactlin}), 
the only vector which has a 
nontrivial projection along 
$A_1^{\otimes n}$ is $W_T^{\otimes n}\ket{T}$. 
Hence, $\alpha_T=0$ leading to a contradiction.
\end{proof}

\begin{claim}
\label{claim:exactquantspan}
$\{W_S^{\otimes n}\}_{S \in {{\bf U} \choose \leq n}}$ 
lie in a vector space of dimension
at most $\sum_{i=0}^{nt}{s \choose i}$.
\end{claim}
\begin{proof}
By definition, for any set $S$, $|S| \leq n$, 
\[ W_S \defeq U_tO_SU_{t-1}O_{S}U_{t-2}O_S \cdots U_2O_SU_1O_SU_0 \]
where $U_0,\ldots,U_{t}$ are unitary transformations (matrices)
independent 
of the set stored. 

For any pair of indices $(i,j)$, 
\begin{eqnarray*}
(W_S)_{i,j} & = & \!\!\!\!\raisebox{-2.5ex}
	           {
		    \(
                     \begin{array}{ll}
                       {\displaystyle \sum_{k_{t-1},\ldots,k_0}}
                        & (U_t)_{i,k_{t-1}} (O_S)_{k_{t-1},k_{t-1}}
		          (U_{t-1})_{k_{t-1},k_{t-2}} 
		          (O_S)_{k_{t-2},k_{t-2}}  \\
			& \cdots (U_1)_{k_1,k_0}
			         (O_S)_{k_0,k_0}(U_0)_{k_0,j} 
                     \end{array}
                    \)
		   }                          \\
            & = & \!\!\!\!\raisebox{-2.5ex}
	           {
		    \(
                     \begin{array}{ll}
                       {\displaystyle \sum_{k_{t-1},\ldots,k_0}}
	                & (U_t)_{i,k_{t-1}} (-1)^{[x(S)]_{l_{k_{t-1}}}}
	                  (U_{t-1})_{k_{t-1},k_{t-2}}
	                  (-1)^{[x(S)]_{l_{k_{t-2}}}} \\
	                & \cdots (U_1)_{k_1,k_0}(-1)^{[x(S)]_{l_{k_0}}}
		                 (U_0)_{k_0,j}
                     \end{array}
                    \)
		   }
\end{eqnarray*}
where, recalling the notation of Section~\ref{sec:defnot}, $x(S)$ is 
the string stored by the storage scheme for set $S$ and $l_i$ is 
either the single location in the string corresponding to index $i$ 
or the empty set.

Therefore, if we define $T_{k_{t-1},\ldots,k_0} \defeq
l_{k_{t-1}} \bigtriangleup l_{k_{t-2}} \bigtriangleup \cdots 
\bigtriangleup l_{k_0}$ and $[x(S)]_T$ to be 
the parity of the bits stored in $x(S)$ at the locations of $T$, we have
% A4wide version
\begin{eqnarray*}
(W_S)_{i,j} & = &\sum_{k_{t-1},\ldots,k_0} 
(-1)^{[x(S)]_{T_{k_{t-1},\ldots,k_0}}}
(U_t)_{i,k_{t-1}}(U_{t-1})_{k_{t-1},k_{t-2}}
       \cdots (U_1)_{k_1,k_0}(U_0)_{k_0,j}\\
& = & \sum_{T \in {[s] \choose \leq t}} (-1)^{[x(S)]_T} 
\sum_{\stackrel{k_{t-1},\ldots,k_0}{T_{k_{t-1},\ldots,k_0}=T}} 
(U_{t})_{i,k_{t-1}} (U_{t-1})_{k_{t-1},k_{t-2}} 
       \cdots (U_1)_{k_1,k_0} (U_0)_{k_0,j}
\end{eqnarray*}
% ordinary version
%\begin{eqnarray*}
%(W_S)_{i,j} & = &\sum_{k_{t-1},\ldots,k_0} 
%(-1)^{[x(S)]_{T_{k_{t-1},\ldots,k_0}}}
%\left( \stackrel{(U_t)_{i,k_{t-1}}(U_{t-1})_{k_{t-1},k_{t-2}}}
%                {\cdots (U_1)_{k_1,k_0}(U_0)_{k_0,j}}
%\right)\\
%& = & \sum_{T \in {[s] \choose \leq t}} (-1)^{[x(S)]_T} 
%\sum_{\stackrel{k_{t-1},\ldots,k_0}{T_{k_{t-1},\ldots,k_0}=T}} 
%\left( \stackrel{(U_{t})_{i,k_{t-1}} (U_{t-1})_{k_{t-1},k_{t-2}}}
%                {\cdots (U_1)_{k_1,k_0} (U_0)_{k_0,j}}
%\right)
%\end{eqnarray*}

Let us define for every set $T \subseteq [s]$, 
$|T| \leq t$, a matrix $A_T$ as 
follows: 
\[ (A_T)_{i,j} \defeq 
\sum_{\stackrel{k_{t-1},\ldots,k_0}{T_{k_{t-1},\ldots,k_0}=T}} 
(U_{t})_{i,k_{t-1}} (U_{t-1})_{k_{t-1},k_{t-2}} 
       \cdots (U_1)_{k_1,k_0} (U_0)_{k_0,j}
\]
Then we have,
\begin{equation}
\label{eq:fourierexpansion}
W_S=\sum_{T \in {[s] \choose \leq t}} (-1)^{[x(S)]_T} A_T
\end{equation}
Hence,
\begin{eqnarray*}
(W_S)^{\otimes n} & = &\left(\sum_{T_{1} \in {[s] \choose \leq t}} 
(-1)^{[x(S)]_{T_1}} A_{T_1}\right) \otimes \cdots \otimes 
\left(\sum_{T_{n} \in {[s] \choose \leq t}} 
             (-1)^{[x(S)]_{T_n}} A_{T_n}\right)\\
& = & \sum_{\stackrel{T_{i} \in {[s] \choose \leq t}}{1 \leq i \leq n}}
(-1)^{[x(S)]_{T_1}}\cdots (-1)^{[x(S)]_{T_n}}(A_{T_1} 
         \otimes \cdots \otimes A_{T_n})\\
& = & \sum_{\tilde{T} \in {[s] \choose \leq nt}} 
                (-1)^{[x(S)]_{\tilde{T}}}B_{\tilde{T}}
\end{eqnarray*}
where for $\tilde{T} \in {[s] \choose \leq nt}$,
\begin{displaymath}
B_{\tilde{T}} \defeq
   \sum_{\stackrel{T_1 \bigtriangleup \cdots \bigtriangleup T_n
                       = \tilde{T}
		  }
		  {T_{i} \in {[s] \choose \leq t}, 1 \leq i \leq n
		  }
        } A_{T_1} \otimes \cdots \otimes A_{T_n}
\end{displaymath}

Hence, we see that 
$\{B_{\tilde{T}}\}_{\tilde{T} \in {[s] \choose \leq nt}}$
span $\{W_S^{\otimes n}\}_{S \in {{\bf U} \choose \leq n}}$. So, 
$\{W_S^{\otimes n}\}_{S \in {{\bf U} \choose \leq n}}$ lie 
in a vector space of 
dimension at most $\sum_{i=0}^{nt} {s \choose i}$. 
\end{proof}

Now the theorem is an easy consequence of the above two claims.
\end{proof}

\smallskip
\noindent {\bf Remark:} Equation~\ref{eq:fourierexpansion} in the
proof of Claim~\ref{claim:exactquantspan} above is 
similar to the statement of a lemma of Shi.
\begin{lemma}[\mbox{\cite[Lemma 2.4]{shi:influence}} rephrased]
Consider a quantum query algorithm with
initial state vector $\kettle{0}$,
with the black
box unitary transformation representing a bit string 
$x = x_1, \ldots, x_s$.
Let $\kettle{\phi}$ be the state vector of the circuit after
$t$ queries to the black box. Then
\begin{displaymath}
\kettle{\phi} = \sum_{T \in {[s] \choose \leq t}} \hat{\phi}_T 
                                                  (-1)^{[x]_T}
\end{displaymath}
where the $\hat{\phi}_T$ are independent of $x$.
\end{lemma}
Shi proved his lemma
using the observation by Beals 
{\em et al.} (see \cite[Lemma 4.1]{beals:quantpoly})
that the amplitudes of the basis states in the
state vector $\kettle{\phi}$ are multilinear polynomials of degree
at most $t$ in $x_1, \ldots, x_s$.

The space-time tradeoff equation for the exact quantum
case holds for the one-sided error case too, as shown below.
% theorem for s-t tradeoff in quantum +ve error
\begin{theorem}
\label{thm:onesidedquant}
The tradeoff result  of Theorem~\ref{thm:exactquant} 
also holds for a quantum
scheme where the query scheme may err with probability less than $1$
on the positive instances
(i.e. if an element is present it may be erroneously reported 
absent), but not on the negative instances
(i.e. if an element is absent it has to be reported absent).
\end{theorem}
\begin{proof} {\bf (Sketch)}
Essentially, the same proof of 
Theorem~\ref{thm:exactquant} goes through.
Since the query scheme can make an error only if the element is 
present,
we observe that the only vector in the linear 
combination (equation~\ref{eq:exactlin}) that has a non-zero projection 
on the
space $A_{1}^{\otimes n}$, is the vector $W_{T}^{\otimes n} \ket{T}$.
Hence $\alpha_{T} = 0$, and the operators 
$\left\{W_{S}\right\}_{S \subseteq {\bf U}, |S| \leq n}$ continue to
be linearly independent. Hence, the same tradeoff
equation holds in this case too.
\end{proof}

We now prove the lower bound on the space used by a two-sided
$\epsilon$-error 1-probe quantum scheme.

\begin{theorem}
\label{thm:twosidedquantone}
Let $n/m < \epsilon < 1/8$.
Suppose there is a scheme 
which stores subsets $S$ of size at most $n$ from a universe
${\mathbf U}$ of size $m$
that answers membership queries, with two-sided error at most 
$\epsilon$, using one quantum probe. It must
use space 
\[ s= \Omega \left(\frac{n\log (m/n)}{\epsilon^{1/6}\log (1/\epsilon)}
             \right)
\]
\end{theorem}
\begin{proof}
Since we are looking at a one probe quantum scheme, $W_S=U_1 O_S U_0$. 
We start
by picking a family $F$ of sets, $F=\{S_1,\ldots,S_k\}$, $S_i \subseteq
{\bf U}$, 
$|S_i|=n$ and $|S_i \cap S_j| \leq n/2$ for all $i \neq j$. By picking
the sets greedily~\cite{erdos:set,nisan:hardness}, one 
obtains a family $F$ with 
\begin{equation}
\label{eq:nwlb}
|F| \geq \frac{{m \choose n}}
              {{n \choose \frac{n}{2}} 
               {{m -\frac{n}{2}} \choose \frac{n}{2}}
              } 
    \geq \frac{\frac{m}{n} \frac{m-1}{n-1} \cdots 
               \frac{m - n/2 + 1}{n - n/2 + 1}
              }
              {2^n}
    \geq \frac{\left(\frac{m}{n}\right)^{n/2}}{2^n}
      =  \left(\frac{m}{4n}\right)^{n/2}
\end{equation}
Let $t \defeq \ceil{\frac{4\log |F|}{n \log (1/(4\epsilon))}}$. 
Since, $m/n \geq 1/\epsilon$,
\begin{displaymath}
\frac{4\log |F|}{n \log (1/(4\epsilon))} 
    \geq \frac{4n \log (m/(4n)}{2n \log (1/(4\epsilon))}
    \geq 2
\end{displaymath}
Hence,
\begin{equation}
\label{eq:quantonetub}
\frac{4\log |F|}{n \log (1/(4\epsilon))} \leq t \leq 
\frac{4\log |F|}{n \log (1/(4\epsilon))} + 1 \leq
\frac{6\log |F|}{n \log (1/(4\epsilon))}
\end{equation}
\begin{claim}
\label{claim:twosidedquantonelin}
$\{W_S^{\otimes nt}\}_{S \in F}$ are linearly independent.
\end{claim}
\begin{proof}
Suppose there is a non-trivial linear combination
\[ \sum_{S \in F} \alpha_S W_S^{\otimes nt} = 0 \]
Fix a $T \in F$. Let $T=\{i_1,\ldots,i_n\}$. Define
\[ \ket{T} \defeq (\ket{i_1} \otimes \ket{i_2} \otimes \cdots \otimes 
\ket{i_n})^{\otimes t}\]
Applying $\phi_T$ to the above linear combination, we get 
\[ \sum_{S \in F} \alpha_S W_S^{\otimes nt}\ket{T} = 0 \]
\[ \Rightarrow \sum_{S \in F} \alpha_S (W_S\ket{i_1}\otimes \cdots
\otimes W_S\ket{i_n})^{\otimes t} = 0 \]
Taking inner product of the above combination with the vector
\[W_T^{\otimes nt}\ket{T}=(W_T\ket{i_1}\otimes \cdots \otimes 
W_T\ket{i_n})^{\otimes t}\]
we get
\begin{displaymath}
\sum_{S \in F} \alpha_{S} 
\langle(W_S\ket{i_1} \otimes \cdots \otimes W_S\ket{i_n})^{\otimes t} |  
       (W_T\ket{i_1} \otimes \cdots \otimes W_T\ket{i_n})^{\otimes t} 
   \rangle =  0  
\end{displaymath}
\begin{equation}
\label{eq:twosidedlin}
\Rightarrow \sum_{S \in F} \alpha_{S} 
                  ( \braket{i_1}{i_1} \cdots \braket{i_n}{i_n})^t = 0 
\end{equation}
\begin{itemize}
\item
For any $i_j \in S \cap T$, $|\braket{i_j}{i_j}| \leq 1$.
\item
For any $i_j \in T$, $W_T\ket{i_j}=v_0 + v_1$ 
where $1 \geq \|v_1\| \geq \sqrt{1-\epsilon}$ and
$\|v_0\|\leq \sqrt{\epsilon}$, $v_0 \in A_0$ and $v_1 \in A_1$. 
For any $i_j \in T-S$, $W_S\ket{i_j}=u_0 + u_1$
where $1 \geq \|u_0\| \geq \sqrt{1-\epsilon}$ and
$\|u_1\|\leq \sqrt{\epsilon}$ and $u_0 \in A_0$ and $u_1 \in A_1$. 
Hence
\begin{eqnarray*}
|\braket{i_j}{i_j}| & =    & |\inprod{u_0}{v_0} + \inprod{u_1}{v_1}| \\
                    & \leq & \|u_0\|\|v_0\| + \|u_1\|\|v_1\| \\
                    & \leq & 2\sqrt{\epsilon} \defeq \delta
\end{eqnarray*}
\end{itemize}

We now note that for every $T \in F$, we have a linear combination
as in equation~\ref{eq:twosidedlin} above.   We can write the 
linear combinations in the matrix form
as $\alpha M = 0$, where $\alpha = (\alpha_S)_{S \in F}$ and $M$
is a $|F| \times |F|$ matrix whose rows and columns are indexed 
by members of $F$. For $S,T \in F$, 
\begin{displaymath}
M(S,T) = (\braket{i_1}{i_1} \cdots \braket{i_n}{i_n})^t
\end{displaymath}
where $T=\{i_1,\ldots,i_n\}$.
The diagonal entries of $M$, $M(T,T)$, are $1$. The 
non-diagonal entries satisfy $|M(S,T)| \leq (\delta)^{(n-|S\cap T|)t}
\leq \delta^{nt/2}$.

Using the lower bound on $t$ from (\ref{eq:quantonetub}), we get
\begin{displaymath}
|F| \delta^{tn/2} = |F| (4\epsilon)^{tn/4}
               \leq 1
\end{displaymath}
Hence $(|F| - 1) (\delta)^{tn/2} < 1$.
This implies that $M$ is non-singular. [Suppose not. Let $y$ be a vector
such that $My = 0$. Let $i$ be the location of the 
largest coordinate of $y$.
We can assume without loss of generality that $y_i = 1$. Now, the $i$th
coordinate of the vector $My$ is at 
least $1 - (|F| - 1) (\delta)^{tn/2} > 0$
in absolute value, which is a contradiction.]
So, $\alpha_S=0$ for all $S \in F$. Hence 
$\{W_S^{\otimes nt}\}_{S \in F}$ are linearly independent.
\end{proof}

\begin{claim}
\label{claim:twosidedquantonespan}
$\left\{W_{S}^{\otimes nt}\right\}_{S \in F}$ lie in a vector space
of dimension at most $\sum_{j=0}^{nt} {s \choose j}$.
\end{claim}
\begin{proof}
Similar to proof of Claim~\ref{claim:exactquantspan} in 
Theorem~\ref{thm:exactquant}.
\end{proof}

Using the two claims above,
\begin{displaymath}
|F| \leq \sum_{j=0}^{nt} {s \choose j} \leq {{s+nt} \choose {nt}}
    \leq \left(\frac{(s+nt)e}{nt}\right)^{nt}
\end{displaymath}
Using the upper bound on $t$ from (\ref{eq:quantonetub}), we  get
\begin{displaymath}
\left(\frac{1}{4\epsilon}\right)^{nt/6} \leq |F| \leq
         \left(\frac{(s+nt)e}{nt}\right)^{nt}
\end{displaymath}
\begin{displaymath}
\Rightarrow
s \geq \frac{nt}{e} \left(\left(\frac{1}{4\epsilon}\right)^{1/6}
                          - e
		    \right)
\end{displaymath}

For values of $\epsilon$ such that $(1/4\epsilon)^{1/6} > 2e$, that is
$\epsilon < 4^{-1} (2e)^{-6}$, using
(\ref{eq:nwlb}) and the lower bound on 
$t$ from (\ref{eq:quantonetub}), we get
\begin{displaymath}
s \geq \frac{nt}{2e} \left(\frac{1}{4\epsilon}\right)^{1/6}
  \geq \frac{2 \log |F|}{e (4\epsilon)^{1/6} \log (1/4\epsilon)}
  \geq \frac{n \log (m/4n)}{e (4\epsilon)^{1/6} \log (1/4\epsilon)}
\end{displaymath}
\begin{displaymath}
\Rightarrow s = \Omega 
  \left(\frac{n \log (m/n)}{\epsilon^{1/6} \log (1 / \epsilon)} \right)
\end{displaymath}
For $4^{-1} (2e)^{-6} \leq \epsilon < 1/8$, we recall
the fact that 
$\Omega (n \log (m/n))$ is always a lower bound (the
information-theoretic lower bound) for the storage
space. Thus, for these values of $\epsilon$ too
\begin{displaymath}
s = \Omega 
  \left(\frac{n \log(m/n)}{\epsilon^{1/6} \log (1 / \epsilon)} \right)
\end{displaymath}

Hence, the theorem is proved.
\end{proof}

% 2-sided error, p bit probes, quantum
We now show how to extend the above argument for 2-sided 
$\epsilon$-error quantum
schemes which make $p$ probes.
\begin{theorem}
\label{thm:twosidedquantp}
For any $p \ge 1$ and $n/m < \epsilon < 2^{-3p}$,
suppose there is a scheme 
which stores subsets $S$ of size at most $n$ from a universe
${\mathbf U}$ of size $m$
that answers membership queries, with two-sided error at most
$\epsilon$, using $p$ quantum probes. 
Define $\delta \defeq \epsilon^{1/p}$. The scheme must use space 
\[ s= \Omega\left(\frac{n\log (m/n)}
                  {\delta^{1/6} \log (1/\delta)}
       \right)\]
\end{theorem}
\begin{proof} {\bf (Sketch)}
The proof of this theorem is similar to the proof of 
Theorem~\ref{thm:twosidedquantone}. 
Pick a family $F$ of sets, $F=\{S_1,\ldots,S_k\}$, $S_i \subseteq
{\bf U}$, 
$|S_i|=n$, $|S_i \cap S_j| \leq n/2$ for all $i \neq j$, such that
$|F| \geq (m/4n)^{n/2}$. 
One can prove that $\left\{W_{S}^{\otimes nt}\right\}_{S \in F}$,
$t \defeq \ceil{\frac{4\log |F|}{n \log (1/(4\epsilon))}}$,
are linearly independent in exactly the same fashion as 
Claim~\ref{claim:twosidedquantonelin} in 
Theorem~\ref{thm:twosidedquantone} was proved. 
The difference is that
$\left\{W_{S}^{\otimes nt}\right\}_{S \in F}$ lie in a vector space
of dimension at most $\sum_{j=0}^{pnt} {s \choose j}$ instead of
$\sum_{j=0}^{nt} {s \choose j}$.  This statement can be proved
just as Claim~\ref{claim:exactquantspan} in
Theorem~\ref{thm:exactquant} was proved.
Therefore, by a argument similar to that at the end of the proof
of Theorem~\ref{thm:twosidedquantone}, we get a lower bound 
\begin{displaymath}
\Omega \left( \frac{n \log (m/n)}
                   {\delta^{1/6} \log (1 / \delta)}
       \right )
\end{displaymath}
\end{proof}

\subsection{Classical schemes}
We now give the proof for the space-time tradeoff equation in the
classical deterministic case.

\begin{theorem} 
\label{thm:classicaldet}
Suppose there exists a classical deterministic scheme for
storing subsets $S$ of size at most $n$ from a universe 
${\mathbf U}$ of size $m$
which uses $s$ bits of storage and answers membership queries
with $t$ classical bit probes. Then, 
\begin{displaymath}
\sum_{i=0}^{n} {m \choose i} \leq \sum_{i=0}^{nt} {s \choose i}
\end{displaymath}
\end{theorem}
\begin{proof}
For $1 \leq i \leq m$,
let $f_{i} : \{0,1\}^{s} \rightarrow \reals$ denote
the function for query $i$, which maps bit strings of length $s$ to
$\{0,1\} \subset \reals$  i.e. $f_{i}$ maps 
$x \in \{0,1\}^{s}$ to $1$ iff the query scheme given query $i$ and
bit string $x$ evaluates to $1$.
Consider a mapping 
$\Phi : {{\bf U} \choose \leq n} \rightarrow 
                  (\{0,1\}^{s} \rightarrow \reals )$ i.e. $\Phi$ takes
a subset of the universe of size at most $n$ to a function from
bit strings of length $s$ to the reals. $\Phi$ is defined as follows
\begin{displaymath}
\Phi(\{\}) \defeq {\rm constant \  1 \ function}
\end{displaymath}
\begin{displaymath}
\Phi(S) \defeq f_{i_1} f_{i_2} \cdots f_{i_k}, 
           \ \ S=\{i_1, \cdots i_k\}, \ S \ne \{ \}
\end{displaymath}

\begin{claim}
$\left\{\Phi(S)\right\}_{S \in {{\bf U} \choose \leq n}}$ are linearly
independent over $\reals$.
\end{claim}
\begin{proof}
Suppose there exists a non-trivial linear combination
\begin{displaymath}
\sum_{S \in {{\bf U} \choose \leq n}} \alpha_{S} \Phi(S) = 0
\end{displaymath}
Pick a set $T$ of smallest cardinality such that $\alpha_{T} \ne 0$.
Let $x(T) \in \{0,1\}^{s}$ be the string stored by the storage scheme.
Applying $x(T)$ to the above linear combination, we get
\begin{displaymath}
\sum_{S \in {{\bf U} \choose \leq n}} \alpha_{S} \Phi(S) x(T) = 0
\end{displaymath}
If $S \ne T$, there exists an element $i \in {\bf U}$ such that
$i \in S - T$. Then, $f_{i} (x(T)) = 0$, and hence, $\Phi(S)(x(T)) = 0$.
If $S = T$, then $\Phi(S)(x(T)) = \Phi(T)(x(T)) = 1$. Hence,
$\alpha_{T} = 0$ which is a contradiction. Hence the claim is proved.
\end{proof}

\begin{claim}
$\left\{\Phi(S)\right\}_{S \in {{\bf U} \choose \leq n}}$ lie in
a vector space of dimension
at most $\sum_{i=0}^{nt}{s \choose i}$.
\end{claim}
\begin{proof}
Since the query scheme is deterministic and
makes at most $t$ (classical) bit probes, given a query 
$i$, $1 \leq i \leq m$, the function $f_i$ 
is modelled by a decision tree of depth at most $t$. Hence $f_{i}$
can be represented  over $\reals$ as a sum of products of 
at most $t$ linear functions,
where the linear functions are either $y_{j}$ (representing
the value stored at location $j$ in the bit string) 
or $1 - y_{j}$ (representing the
negation of the value stored at location $j$).  Note that
for any $y \in \{0,1\}^{s}$, at most one of these products evaluates
to $1$. Such a function
can be represented as a multilinear polynomial in $y_1,y_2,\ldots,y_s$ 
of degree at most $t$.
A product of at most $n$ such functions can be represented as a
multilinear polynomial of degree at most $nt$.
Hence, $\left\{\Phi(S)\right\}_{S \in {{\bf U} \choose \leq n}}$ 
lie in the span of at  most $\sum_{i=0}^{nt} {s \choose i}$ functions
from $\{0,1\}^{s}$ to $\reals$. From this,  the claim follows.
\end{proof}

From the above two claims, the theorem follows.
\end{proof}

In fact, the tradeoff result can be extended to the one-sided
error classical case too.

\begin{theorem}
\label{thm:classicalonesided}
The tradeoff result of Theorem~\ref{thm:classicaldet} also holds for
classical schemes where the query scheme may err  with
probability less than $1$ on the positive
instances (i.e. if an element is present it might report it to
be absent), but not on the negative instances
(i.e, if an element is absent it has to be reported as absent).
In fact, the tradeoff result holds for nondeterministic query
schemes too.
\end{theorem}
\begin{proof} {\bf (Sketch)}
A proof very similar to that of Theorem~\ref{thm:classicaldet} goes
through. We just observe that now the query scheme is a 
logical disjunction over a family of deterministic query schemes.
If the query element is present in the 
set stored, there is a decision tree in this family that
outputs 1. If the query element is not present in the set stored,
then all the decision trees output 0. Let us denote by $F_i$ the family
of decision trees corresponding to query element $i$, $1 \leq i 
\leq m$. For any decision tree $D$ in $F_i$, let $g_D
:\{0,1\}^s \rightarrow \{0,1\}$ be the function it evaluates.

Let us now define $f_i \defeq \sum_{D \in F_i} g_D$. Then 
 
\[f_i(x[T])  \left\{ \begin{array}{ll}
                        \geq 1 & \mbox{if $i \in T$}\\
		         = 0 & \mbox{otherwise}
                       \end{array}
               \right. \]
With this choice of $f_i$, the rest of the proof is the same 
as in the deterministic case.
\end{proof}

Now we give a simple proof of the lower bound for the space used
by a classical randomised scheme which answers membership queries
with two-sided error at most $\epsilon$ and uses only one bit probe.

\begin{theorem}
\label{thm:classicaltwosidedone}
Let $1/18 > \epsilon > 1 / m^{1/3}$ and $m^{1/3} > 18n$.
Any classical scheme which stores subsets $S$
of size at most $n$ from a universe ${\mathbf U}$ of size $m$ and 
answers membership queries, with two-sided error at most $\epsilon$,
using one bit probe must use space 
\begin{displaymath}
\Omega \left(\frac{n \log m}{\epsilon^{2/5} \log (1 / \epsilon)}
        \right)
\end{displaymath}
\end{theorem}
\begin{proof}
Suppose there is a classical scheme which stores subsets 
of size $n$ from a universe of size $m$ using $s$ bits of storage,
and answers membership queries
using one bit probe with two-sided error at most $\epsilon$.
Define $k \defeq \ceil{\frac{4 \log (27m)}{3 \log (1/4e\epsilon)}}$.
Since $m^{1/3} > 1 / \epsilon$,
$\frac{4 \log (27m)}{3 \log (1/4e\epsilon)} \geq 4$. Therefore,
\begin{equation}
\label{eq:twosidedclassical}
\frac{4 \log (27m)}{3 \log (1/4e\epsilon)} \leq k \leq
\frac{4 \log (27m)}{3 \log (1/4e\epsilon)} + 1    \leq
\frac{5 \log (27m)}{3 \log (1/4e\epsilon)}
\end{equation}

We repeat the query scheme $k$ times and accept only if
more than $3k/4$ trials accept. Then the probability of making
an error on a positive instance (i.e. the query element 
is present in the set stored) is bounded by
\begin{displaymath}
{k \choose k/4} \epsilon^{k/4} \leq (4e)^{k/4} \epsilon^{k/4}
    = (4 e \epsilon)^{k/4}
\end{displaymath}
The probability of making an error on a negative instance (i.e. the
query element is not present in the set stored) is bounded by
\begin{displaymath}
{k \choose 3k/4} \epsilon^{3k/4}
    \leq \left( \frac{4 e \epsilon}{3} \right)^{3k/4}
    \leq (4 e \epsilon)^{3k/4}
\end{displaymath}
From lower bound on $k$ from (\ref{eq:twosidedclassical}), we get
\begin{displaymath}
\begin{array}{lclcl}
\mbox{Pr[Error on a positive instance]}
         & \leq & (4e\epsilon)^{k/4}
         & \leq & \frac{1}{(27m)^{1/3}} \leq \frac{1}{3n} \\
\mbox{Pr[Error on a negative instance]}
         & \leq & (4e\epsilon)^{3k/4}
         & \leq & \frac{1}{27m} 
\end{array}
\end{displaymath}
Hence, the probability that a random sequence of coin tosses gives 
the wrong 
answer on some query $q \in {\bf U}$ and a particular set
$S$ stored, is at most  
\begin{displaymath}
\frac{1}{3n} \times n + \frac{1}{27m} \times (m - n)
        < \frac{1}{2}
\end{displaymath}

Call a sequence of coin tosses bad for a set $S$, if when $S$ is stored,
there is
one query $q \in {\bf U}$ for which the query scheme with these
coin tosses gives the
wrong answer. Thus, at most half of the coin toss sequences are
bad for a fixed set $S$.
By an averaging argument, there exists a sequence of coin tosses which
is bad for at most half of the sets 
$S \in {{\bf U} \choose n}$. By setting the coin tosses to that 
sequence, we now get a deterministic scheme which answers
membership queries correctly for at least half the sets 
$S \in {{\bf U} \choose n}$, 
and uses $k$
bit probes. From the proof
of Theorem~\ref{thm:classicaldet}, we have that
\begin{displaymath}
\frac{1}{2} {m \choose n} \leq \sum_{i=0}^{nk} {s \choose i}
                          \leq {s+nk \choose nk}
                          \leq \left( \frac{e(s+nk)}{nk} \right)^{nk}
\end{displaymath}
\begin{displaymath}
\Rightarrow \frac{1}{2} \left( \frac{m}{n} \right)^n
               \leq \left( \frac{e(s+nk)}{nk} \right)^{nk}
\end{displaymath}

Using the upper bound on $k$ in (\ref{eq:twosidedclassical}) and the
fact that $m^{1/3} > 18n$, we get
\begin{displaymath}
\left(\left(\frac{1}{4e\epsilon}\right)^{3k/5}\right)^{2n/3}
        \leq \left(27m\right)^{2n/3} 
          =  \left(9 m^{2/3}\right)^n 
        \leq \frac{1}{2} \left(18 m^{2/3}\right)^n 
        \leq \frac{1}{2} \left(\frac{m}{n}\right)^n
\end{displaymath}
\begin{displaymath}
\Rightarrow
\left(\frac{1}{4e\epsilon}\right)^{2nk/5}
               \leq \left( \frac{e(s+nk)}{nk} \right)^{nk}
\end{displaymath}
\begin{displaymath}
\Rightarrow
s \geq \frac{nk}{e} \left(\left(\frac{1}{4e\epsilon}\right)^{2/5}
                          - e
		    \right)
\end{displaymath}
Arguing as in the last part of the proof of 
Theorem~\ref{thm:twosidedquantone},
and recalling that since $m^{1/3} > 18n$, $\Omega(n\log m)$ is
always a lower bound (the information-theoretic lower bound)
for the storage space, we get
\begin{displaymath}
s = \Omega \left(\frac{n \log m}{\epsilon^{2/5} \log (1 / \epsilon)}
        \right)
\end{displaymath}
\end{proof}
 
We can extend the classical randomised two-sided error space 
lower bound above to the case of
multiple bit probes.

\begin{theorem}
\label{thm:classicaltwosidedp}
Let $p \ge 1$, $18^{-p} > \epsilon > 1 / m^{1/3}$ and 
$m^{1/3} > 18n$.
Define $\delta \defeq \epsilon^{1/p}$.
Any classical scheme which stores subsets $S$
of size at most $n$ from a universe ${\mathbf U}$ of size $m$ and 
answers membership queries, with two-sided error at most $\epsilon$,
using at most $p$ bit probes must use space 
\begin{displaymath}
\Omega \left(\frac{n \log m}{\delta^{2/5} \log (1 / \delta)}
        \right)
\end{displaymath}
\end{theorem}
\begin{proof} {\bf (Sketch)}
The proof of this theorem is similar to the proof of 
Theorem~\ref{thm:classicaltwosidedone} above. 
We repeat the query scheme 
$k \defeq \ceil{\frac{4 \log (27m)}{3 \log (1/4e\epsilon)}}$ 
times and accept only if more than $3k/4$ trials accept.
We ``derandomise'' the new query scheme in a manner similar to
what was done in the proof of Theorem~\ref{thm:classicaltwosidedone}.
We thus get a deterministic query scheme making $kp$ bit probes and
answering membership queries correctly for at least half the sets
$S \in {{\mathbf U} \choose n}$.
The rest of the proof now follows in the same fashion as the 
proof of Theorem~\ref{thm:classicaltwosidedone}. 
\end{proof}

\section{Conclusion and open problems}
\label{sec:conclusion}
In this paper, we introduce the quantum bit probe model and study
the complexity of the static membership problem in this model.
We study the problem in the exact and bounded (one-sided and
two-sided) error versions of the model and give lower bounds which
almost match the corresponding classical upper bounds.
We also give stronger/simplified proofs of lower bounds for the problem
in the classical setting. 

The paper of Buhrman {\em et al.}~\cite{buhrman:bitprobe} 
also considers classical
schemes for the static membership problem where the error is bounded
and restricted only to negative instances (i.e. when the query element
is not a member of the stored set). For such schemes, they give 
almost matching upper and lower bounds. But for negative one-sided
error quantum schemes, we can only prove similar lower bounds as for
two-sided error quantum schemes. Also, we do not know if there are 
negative one-sided error quantum schemes better than the 
classical ones in \cite{buhrman:bitprobe}. Thus there is 
a gap between the upper and lower bounds
here, and resolving it is an open problem. 

The classical bit probe model has been used in the past
to study other 
static data structure problems like, for example, perfect hashing
and element
containment (see e.g.~\cite{pagh:setmemb, miltersen:bitprobe}). The 
complexity of these
problems in the quantum bit probe model is an important open problem.

\subsection*{Acknowledgements} We thank the anonymous referees
for their comments, which have helped us to improve the presentation
of the paper and fix some bugs in the parameters appearing in some
theorems. We also thank them for pointing out the similarities between
Claim~\ref{claim:exactquantspan} in 
Theorem~\ref{thm:exactquant}, and the
observations of Beals {\em et al.}~\cite{beals:quantpoly} and 
Shi~\cite{shi:influence}.

\bibliography{quantset}
\end{document}